\documentstyle[epsfig,aps,prl,twocolumn]{revtex}
\newcommand{\etal}{{\em{et al.}}}
\newcommand{\Vud}{V_{\mathit{ud}}}
\newcommand{\Vus}{V_{\mathit{us}}}
\newcommand{\Vub}{V_{\mathit{ub}}}
\newcommand{\Vcd}{V_{\mathit{cd}}}
\newcommand{\Vcs}{V_{\mathit{cs}}}
\newcommand{\Vcb}{V_{\mathit{cb}}}
\newcommand{\Vtd}{V_{\mathit{td}}}
\newcommand{\Vts}{V_{\mathit{ts}}}
\newcommand{\Vtb}{V_{\mathit{tb}}}
\newcommand{\VUD}{|V_{\mathit{ud}}|}
\newcommand{\VUS}{|V_{\mathit{us}}|}
\newcommand{\VUB}{|V_{\mathit{ub}}|}
\newcommand{\VCD}{|V_{\mathit{cd}}|}
\newcommand{\VCS}{|V_{\mathit{cs}}|}
\newcommand{\VCB}{|V_{\mathit{cb}}|}
\newcommand{\VTD}{|V_{\mathit{td}}|}
\newcommand{\VTS}{|V_{\mathit{ts}}|}
\newcommand{\VTB}{|V_{\mathit{tb}}|}

\newcommand{\mt}{m_{\mathit{t}}}
\newcommand{\mH}{m_{\mathit{H}}}
\newcommand{\mZ}{m_{\mathit{Z}}}
\newcommand{\mW}{m_{\mathit{W}}}
\newcommand{\VUDCHISQ    }{\chi^2_{\mathit{ud}}}   
\newcommand{\VUSCHISQ    }{\chi^2_{\mathit{us}}}
\newcommand{\VCDCHISQ    }{\chi^2_{\mathit{cd}}}
\newcommand{\VCSCHISQ    }{\chi^2_{\mathit{cs}}}
\newcommand{\VUBCHISQ    }{\chi^2_{\mathit{ub}}}
\newcommand{\VCBCHISQ    }{\chi^2_{\mathit{cb}}}
\newcommand{\VUBCBCHISQ  }{\chi^2_{\mathit{ub/cb}}}
\newcommand{\VTDTBCHISQ  }{\chi^2_{\mathit{td*tb}}}

\newcommand{\VTSTBCBCHISQ}{\chi^2_{\mathit{ts*tb/cb}}}
\newcommand{\VTBCHISQ    }{\chi^2_{\mathit{tb}}}   
\begin{document}
%
%
%
\draft
\title{First unitarity-independent determination of the CKM matrix elements \\
       ${\mathbf{V_{td}}}$, ${\mathbf{V_{ts}}}$, and ${\mathbf{V_{tb}}}$ 
       and the implications for unitarity}
\author{John Swain and Lucas Taylor}
\address{Department of Physics, Northeastern University, Boston, Massachusetts 02115, USA}
\date{\today}
\maketitle
\begin{abstract}
The magnitudes of the CKM matrix elements $\Vtd$, $\Vts$, and $\Vtb$ 
are determined for the first time without any assumptions of unitarity.
The implications for the unitarity of the CKM matrix as a whole are discussed.
\end{abstract}
%
%
%
%
\section{Introduction}
The relationship between weak and mass eigenstates of quarks, 
assuming there are three generations, is described by the  
Cabibbo-Kobayashi-Maskawa (CKM) matrix\cite{CKM}  
\begin{equation}
V = 
  \left( \begin{array}{ccc}
      \Vud & \Vus  &  \Vub  \\
      \Vcd & \Vcs  &  \Vcb  \\
      \Vtd & \Vts  &  \Vtb  
   \end{array}  \right).  
   \label{eq:ckm}
\end{equation}
Unless the CKM matrix is {\em{assumed}} to be unitary the bottom row 
elements, $\Vtd$, $\Vts$, and $\Vtb$, are undetermined.
There are, however, a number of measured quantities which are sensitive
to combinations of these elements. 

We use our recent determination of $\VTB$ from electroweak 
corrections to $Z$ decays~\cite{TAYLOR_VTB} to ``unlock'' 
the bottom row of the CKM matrix.
From a combined fit to data from the LEP, SLC, CESR, Tevatron,
and other experiments we determine all the CKM elements, 
independent of unitarity assumptions, and proceed to test unitarity 
of the full CKM matrix for the first time.

\section{Constraints on the CKM elements}
We use experimental results and the corresponding 
theoretical predictions to construct a chisquare which is then minimised 
to obtain all the CKM matrix elements.
The following sections describe individual chisquare constraints 
which are ultimately summed to form the global chisquare.
\subsection{Constraint on $\VTB$}

In a previous paper~\cite{TAYLOR_VTB} we described a new method for the 
determination of $\VTB$ from electroweak loop corrections, 
in particular to the process $Z\rightarrow b\bar{b}$.
We define a chisquare
$\VTBCHISQ(\mZ, \mt, \mH, \alpha_{\mathrm{s}}, \alpha, \Vtb)$ 
for the agreement of the theory with nineteen experimental measurements.
Full details of the input parameter values, errors, and  
correlations are given in Ref.~\cite{TAYLOR_VTB}.
\subsection{Constraint on $|\Vtb^*\Vtd|$}
The product $|\Vtb^*\Vtd|$ is constrained by measurements of 
$B^0_d-{\bar{B}}^0_d$ mixing.
Therefore, using these measurements and our constraint on 
$\VTB$ enables us to extract $\VTD$.
The mixing is expressed in terms of the mass difference 
$\Delta{m}_d$ of the CP eigenstates~\cite{BDECAYS_ROSNER}
\begin{eqnarray}
\Delta{m}_d & = & \frac{G^2_F}
                       {6\pi^2}
                       |\Vtb^*\Vtd|^2     
                       \mW^2
                       m_{B_d}
                       \left(
                       B_{B_{d}}
                       f_{B_d}^2
                       \right) 
                       \eta_B
                       S_0\left(x_t\right) 
                            \label{eq:deltam},
\label{equ:deltamd}
\end{eqnarray}
where
${\sqrt{B_{B_d}}} f_{B_d} = (201 \pm 42)$\,MeV~\cite{FLYNN97A}
comes from a combined analysis of lattice calculations, which are
compatible with predictions of QCD sum rules, and 
the QCD factor is $\eta_B = 0.55\pm0.01$~\cite{BURAS97A}.
The function $S_0(x_t = {\overline{m}}_t^2/m_W^2)$ is given by~\cite{BDECAYS_ROSNER}
\begin{equation} 
S_0(x_t) = \frac{x_t}{4} \left[ 1 + 
                          \frac {3-9x_t}{\left(x_t-1\right)^2}
                          +
                          \frac {6 x_t^2 {\mathrm{ln}} x_t}{\left(x_t-1\right)^3}
                   \right],
\end{equation}
where ${\overline{m}}_t$ is the running top mass~\cite{BURAS97A} as required 
for consistency with $\eta_B$. 
It is related to the pole mass $m_t$ as measured by the Tevatron, according to 
\begin{equation}
{\overline{m}}_t = \mt  \left( 1 - \frac{4}{3} 
                                         \frac{\alpha_s(\mt )}
                                         {\pi}
                                   \right) \nonumber \\
                 \approx \mt - 8\,{\mathrm{GeV}}. 
\end{equation}
Given the measurement of 
$\Delta{m}_d = 0.472 \pm 0.018\,{\mathrm{ps}}^{-1}$~\cite{JIMACK_EPS97}
we define the chisquare constraint on $|\Vtb^*\Vtd|$ 
\begin{equation}
\VTDTBCHISQ = \frac {\left(\Delta{m}_d(\mt,\mW)         - 0.472\right)^2}  
                    {\left(0.418\Delta{m}_d\left(\mt,\mW\right)\right)^2 
                    + (0.018)^2}
\end{equation}
where the theoretical prediction in the numerator depends on $\mt$ and $\mW$.
We vary $\mt$ explicitly in the fit; $\mW$ is varied implicitly since 
it is a function of the other electroweak parameters which vary 
explicitly~\cite{TAYLOR_VTB}.
The two terms in the denominator correspond respectively to the 
theoretical and experimental uncertainties, the former being dominated  
by the 20\% relative error on ${\sqrt{B_{B_d}}} f_{B_d}$. 
\subsection{Constraint on $|\Vts^* \Vtb | / |\Vcb|$}
%
The rate for the process $b\rightarrow s\gamma$, which involves a 
loop with a virtual top quark, is proportional to $|\Vts^*\Vtb|^2$. 
Our constraint on $\VTB$ therefore enables us to determine $\VTS$.
The theoretical prediction in the Standard Model has been 
calculated to be~\cite{CHETYRKIN97A}
\begin{eqnarray}
{\mathrm{BR}} (b\rightarrow s\gamma)_{\mathrm{th.}} & = &
(3.28 \pm 0.33)\times 10^{-4} 
\nonumber \\
& & 
\times 
\left[ 
\frac{ |\Vts^* \Vtb | / | \Vcb | }
     { 0.976 }
\right]^2.
\end{eqnarray}
The element $\Vcb$ enters through the normalisation to the 
measurement of
${\mathrm{BR}} (B \rightarrow X_c e \bar{\nu}_e) 
= (10.4\pm0.4)\%$~\cite{PDG96SHORT} 
which reduces the otherwise large error due to the 
b-quark mass uncertainty.
%
%
CLEO has measured 
${\mathrm{BR}} (b\rightarrow s\gamma) = 
(2.32 \pm 0.57 \pm 0.35)\times 10^{-4}$~\cite{CLEO_BTOSGAMMA},
and ALEPH has presented a preliminary measurement of 
${\mathrm{BR}} (b\rightarrow s\gamma) = 
(3.29 \pm 0.71 \pm 0.68)\times 10^{-4}$~\cite{KLUIT_EPS97}.       
We average these two results, assuming a common uncertainty of 
$0.3\times 10^{-4}$ for the theoretical modelling, to obtain 
${\mathrm{BR}}(b\rightarrow s\gamma)_{\mathrm{ex.}} = 
(2.60 \pm 0.59)\times 10^{-4}$. 
The chisquare constraint on $|\Vts^* \Vtb | / |\Vcb|$ is then 
\begin{equation}
\VTSTBCBCHISQ = \frac {( {\mathrm{BR}} (b\rightarrow s\gamma)_{\mathrm{th.}} - 
                         {\mathrm{BR}} (b\rightarrow s\gamma)_{\mathrm{ex.}})^2}
                      {   \sigma_{\mathrm{th.}}^2 + 
                          \sigma_{\mathrm{ex.}}^2},
\end{equation}
where $\sigma_{\mathrm{th.}}$ and $\sigma_{\mathrm{ex.}}$ are the theoretical 
and experimental errors on ${\mathrm{BR}} (b\rightarrow s\gamma)$ respectively.
We neglect the explicit dependence on $\mt$ and $\alpha_{\mathrm{s}}$ 
but we do include their 3\% contribution to the theoretical 
uncertainty of 10\%.

\subsection{Constraints on other CKM elements}
%
The most precise constraints on the element $\VUD$ are from
nuclear beta decay.
We use the recent measurement of $\VUD$ from the Chalk River 
Laboratories~\cite{HAGBERG96A} to 
define the chisquare constraining $\VUD$
\begin{equation}
\VUDCHISQ = {(\VUD - 0.9740)^2} /
                {(0.0005)^2}.
\end{equation}
%
%
We use combined results from measurements of $K_{e3}$ and hyperon
decays~\cite{PDG96SHORT} to construct the $\VUS$ chisquare  
\begin{equation}
\VUSCHISQ = {(\VUS - 0.2205)^2} /
                {(0.0018)^2}.
\end{equation}
%
%
We use the results of experiments measuring the production 
of charm from (anti-)neutrino interactions with $d$ valence 
quarks~\cite{PDG96SHORT} to construct the 
$\VCD$ chisquare 
\begin{equation}
\VCDCHISQ = {(\VCD - 0.224)^2} /
                {(0.016)^2}.
\end{equation}
%
%
Measurements of the $D_{e3}$ decay 
$D\rightarrow\bar{K}e^+\bar{\nu}$~\cite{PDG96SHORT} are used to construct 
the $\VCS$ chisquare 
\begin{equation}
\VCSCHISQ = {(\VCS - 1.01)^2} /
                {(0.18)^2},
\end{equation}
where the error is dominated by uncertainties in the hadronic form factors.
 
The existence of $b\rightarrow u$ transitions, which depend on 
$\VUB$, is well established from measurements of 
the endpoint of the charged lepton energy spectrum in 
$b\rightarrow X\ell^-\bar{\nu}$ decays.
Such analyses yield the ratio $|\Vub| / |\Vcb|$~\cite{PDG96SHORT}
from which we derive the chisquare
\begin{equation}
\VUBCBCHISQ = {(|\Vub| / |\Vcb| - 0.08)^2} /
                  {(0.02)^2}.
\end{equation}
%
%
Recently CLEO reported the first measurements of 
${\mathrm{BR}}\left(B^0 \rightarrow \pi^-\ell^+\nu\right) =
\left(1.8 \pm 0.4 \pm 0.3 \pm 0.2\right)\times 10^{-4}$
and 
${\mathrm{BR}}\left(B^0 \rightarrow \rho^-\ell^+\nu\right) =
\left(2.5 \pm 0.4 ^{+0.5}_{-0.7} \pm 0.5\right)\times 10^{-4}$
where the errors correspond respectively to the statistical, 
systematic, and model-dependent uncertainties~\cite{CLEO_VUB_EXCL}.
These results yield the chisquare
\begin{equation}
\VUBCHISQ = {(\VUB - 0.0033)^2} /
            {(0.0008)^2}.
\end{equation}
%
%
%
The element $\VCB$ is determined both from the average b-lifetime 
and from $B\rightarrow D^*\ell\nu$ decays in the limit 
of zero recoil.
The resulting chisquare for $\VCB$ is~\cite{BURAS97A} 
\begin{equation}
\VCBCHISQ = {(\VCB - 0.040)^2} /
                {(0.003)^2}.
\end{equation}
%
\section{Unitarity-free fit}
We fit for all the CKM matrix elements without assuming 
unitarity of the CKM matrix by minimising the chisquare, 
defined as 
\begin{eqnarray}
\chi^2 
       & = &   \VUDCHISQ       (\Vud)
	     + \VUSCHISQ       (\Vus)
	     + \VCDCHISQ       (\Vcd)
	     + \VCSCHISQ       (\Vcs)                                     
	       \nonumber \\
       &   &      
	     + \VUBCBCHISQ     (\Vub, \Vcb)
	     + \VUBCHISQ       (\Vub)
	     + \VCBCHISQ       (\Vcb)
	       \nonumber \\
       &   &      
	     + \VTDTBCHISQ     (\Vtb, \Vtd, \mt, \mW)
	       \nonumber \\
       &   &      
	     + \VTSTBCBCHISQ   (\Vts, \Vtb, \Vcb)
	       \nonumber \\
       &   &      
	     + \VTBCHISQ       (\mZ, \mt, \mH, \alpha_{\mathrm{s}}, \alpha, \Vtb),  
\end{eqnarray}
where for completeness we have included all the dependences of the 
individual chisquare definitions.
The 14 parameters which vary in the fit are 
$\mZ$,                   
$\mt$,                 
$\log_{10}(\mH)$,
$\alpha_{\mathrm{s}}(m_Z)$,     
$\alpha$, and 
the magnitudes of the nine CKM elements.
The results of the fit are shown in the second column 
of table~\ref{tab:results}, where the errors include the 
effects of experimental and theoretical 
uncertainties and all correlations.
The chisquare per degree of freedom for the unitarity-free fit is 
$15.1 / (28 - 14) = 15.1 / 14$.
The probability to obtain a value greater than this is 37\% indicating 
that the data are in good agreement with the underlying model.

The bottom row CKM elements, $\VTD$, $\VTS$, and $\VTB$ are determined 
for the first time, independent of any unitarity assumptions. 
\section{Unitarity-constrained fit}
We also fit the data imposing CKM unitarity as a constraint.
The four parameters which define a three dimensional unitary matrix
can be uniquely determined from the moduli of the elements. 
We choose the parametrisation of the CKM matrix advocated 
by the Particle Data Group\cite{PDG96SHORT} which enforces exact
unitarity 
\begin{eqnarray}
  \Vud & = &     c_{12} c_{13}                                                \\
  \Vus & = &     s_{12} c_{13}                                                \\
  \Vub & = &     s_{13}                               e^{-i\delta_{13}}       \\
  \Vcd & = &   - s_{12} c_{23} - c_{12} s_{23} s_{13} e^{i\delta_{13}}        \\
  \Vcs & = &     c_{12} c_{23} - s_{12} s_{23} s_{13} e^{i\delta_{13}}        \\
  \Vcb & = &     s_{23} c_{13}                                                \\
  \Vtd & = &     s_{12} s_{23} - c_{12} c_{23} s_{13} e^{i\delta_{13}}        \\
  \Vts & = &   - c_{12} s_{23} - s_{12} c_{23} s_{13} e^{i\delta_{13}}        \\
  \Vtb & = &     c_{23} c_{13}                                                
  \label{eq:ckmfit}
\end{eqnarray}
where 
$c_{ij} = \cos\theta_{ij}$ and  
$s_{ij} = \sin\theta_{ij}$ for three angles, 
$\theta_{12}$, 
$\theta_{13}$, and 
$\theta_{23}$, which lie in the first quadrant.
The phase parameter $\delta_{13}$ lies in the range 
$0 < \delta_{13} < 2\pi$. 
The 9 parameters which vary in the unitarity-constrained fit are 
$\mZ$,                   
$\mt$,                 
$\log_{10}(\mH)$,
$\alpha_{\mathrm{s}}(m_Z)$,     
$\alpha$, 
$s_{12}$, $s_{13}$, $s_{23}$, and $\cos\delta_{13}$. 
The results of the fit are shown in table~\ref{tab:results} together
with the corresponding values of the CKM elements.

The changes in $\mt$, $\mH$, and $\alpha_{\mathrm{s}}$, compared to
the unitarity-free fit, are due to their
correlation with $\VTB$, as discussed in Ref.~\cite{TAYLOR_VTB}.
Since only terms of the form $\cos\delta_{13}$ appear 
in this analysis, there is a two-fold ambiguity between the regions 
$0 < \delta_{13} < \pi$, as favoured by CP non-conservation in the 
kaon system, and $\pi < \delta_{13} < 2\pi$.         
Unfortunately, the fitted value of $\cos\delta_{13} = -0.26^{+0.82}_{-0.74}$ 
has rather large errors; in fact, the lower error 
corresponds to the physical limit of $\cos\delta_{13} = -1$.

The chisquare per degree of freedom for the unitarity-constrained fit is 
$25.5 / (28 - 9)  = 25.5 / 19$.
The probability to obtain a value larger than this is 14\% 
which indicates reasonable consistency of the data and the model, 
albeit with less probability than the unitarity-free fit.

\section{Tests of CKM unitarity}
Unitarity implies $V V^\dagger = V^\dagger V = U$, where $U$ is the unit matrix, 
and hence the normality conditions  
\begin{eqnarray}
\rho_i   \equiv |V_{i1}|^2 + |V_{i2}|^2 + |V_{i3}|^2 & = & 1;\mbox{~~~}{\mathrm{and}}  \\
\kappa_j \equiv |V_{1j}|^2 + |V_{2j}|^2 + |V_{3j}|^2 & = & 1;   
\end{eqnarray}
for $i=1,2,3$ rows and $j=1,2,3$ columns.
From only the magnitudes of the nine CKM elements it is not
possible to test orthogonality of pairs of different rows 
or pairs of different columns.
Allowing for correlations, the results of the six normality tests are
\begin{eqnarray}
\rho_1   \equiv \VUD^2 + \VUS^2 + \VUB^2 & = &  0.997 \pm 0.001; \label{eq:row1} \\
\rho_2   \equiv \VCD^2 + \VCS^2 + \VCB^2 & = &  1.105 \pm 0.367; \label{eq:row2} \\
\rho_3   \equiv \VTD^2 + \VTS^2 + \VTB^2 & = &  0.623 \pm 0.315; \label{eq:row3} \\
\kappa_1 \equiv \VUD^2 + \VCD^2 + \VTD^2 & = &  0.999 \pm 0.007; \label{eq:col1} \\
\kappa_2 \equiv \VUS^2 + \VCS^2 + \VTS^2 & = &  1.104 \pm 0.367; \label{eq:col2} \\
\kappa_3 \equiv \VUB^2 + \VCB^2 + \VTB^2 & = &  0.629 \pm 0.316. \label{eq:col3}
\end{eqnarray}
The respective similarities between the three row constraints and the 
three column constraints reflects the relatively small values of the 
off-diagonal elements.
Only $\rho_1$ is significantly different from unity, being $2.1$ standard 
deviations low.
Given that there are six, albeit somewhat correlated, conditions we do not 
attach great significance to this although its future evolution is
of interest.

The trend is for the $\rho_i$ and $\kappa_j$ to be somewhat lower 
than the expectations of a unitary matrix, as would be the case 
if there were more than three generations. 
To test this possibility with a single number we define the quantity
\begin{equation}
\Delta_{\mathrm{CKM}} = \sqrt{(\rho_1\rho_2\rho_3)(\kappa_1\kappa_2\kappa_3)}
\end{equation}
which should be unity for a unitary matrix.
We find $\Delta_{\mathrm{CKM}} = 0.69 \pm 0.43$ where all 
correlations have been taken into account. 
This is lower than unity but only by $0.7$ standard deviations, 
indicating that there is no compelling evidence for 
a deviation from unitarity. 

These results constitute the first tests of unitarity of 
the complete CKM matrix.
In particular, $\rho_3$, $\kappa_1$, $\kappa_2$, $\kappa_3$, 
and $\Delta_{\mathrm{CKM}}$ are measured for the first time.

\section{Outlook}
The error on $\VTD$ is limited by the large theoretical 
uncertainty on $(B_{B_d} f_{B_d}^2)$, which is improving 
only slowly with time. 
In future $\VTD$ will also be determined from the 
process $K^+ \rightarrow \pi^+\nu\bar{\nu}$, which 
is sensitive to the product $|\Vtd^*\Vts|$.
Recently, the E787 experiment observed one candidate event with 
an estimated background of $(0.08 \pm 0.03)$ events, from which 
they determined
${\mathrm{BR}}(K^+ \rightarrow \pi^+ \nu\ \bar{\nu})
= (4.2^{+9.7}_{-3.5})\times 10^{-10}$~\cite{KTOPINUNU}.
From this they derive, assuming unitarity, the constraint 
$0.006 < \VTD < 0.060$, which is in agreement with our 
more precise unitarity-independent result.
Ultimately, the measurement of 
${\mathrm{BR}}(K^+ \rightarrow \pi^+ \nu\ \bar{\nu})$ will 
yield precise and theoretically well understood constraints 
on $|\Vtd^*\Vts|$. 

The uncertainty on $\VTS$ from the process $b\rightarrow s\gamma$ 
is statistics limited, so some improvement can be 
expected from CLEO and the B factory experiments.
The parameter $\Delta m_s$ describing $B_s-\bar{B_s}$ mixing is 
also sensitive to $\VTS$.
The latest experimental limit is 
$\Delta{m}_s > 10.2\,{\mathrm{ps}}^{-1}$ at the 95\% C.L.~\cite{JIMACK_EPS97}.
If $\Delta{m}_s$ were to be measured, its interpretation in terms 
of $\VTS$ would suffer from the large errors on $(B_{B_s} f_{B_s}^2)$.
The ratio $\Delta{m}_d/\Delta{m}_s$ is, however, considerably 
less sensitive to theoretical uncertainties and its measurement 
would yield an important constraint on the ratio $|\Vtd/\Vts|^2$.
Similarly, a future measurement of the ratio
${\mathrm{BR}}(B\rightarrow (\rho/\omega)\gamma) /
 {\mathrm{BR}}(B\rightarrow      K^*     \gamma)$  
would constrain $|\Vtd/\Vts|^2$.  

Given that LEP has finished running on the Z, the uncertainty 
on $\VTB$ from our method of fitting electroweak data is 
unlikely to go below approximately 20\%~\cite{TAYLOR_VTB}.
The single top quark production rate at hadron colliders is, however, 
also sensitive to $\VTB$.
The estimated uncertainty at the end of the Tevatron Run II is 
$\delta\VTB/\VTB = 12\%-19\%$, where the range reflects the 
uncertainty of the gluon structure functions~\cite{HEINSON97B}.

\section{Summary}
The elements $\VTD$, $\VTS$, and $\VTB$, and hence the full 
CKM matrix, are determined for the first time independent of 
any unitarity assumptions: 
\begin{eqnarray}
\VTD & = & 0.0113  _{-0.0029}^{+0.0060};   \\ 
\VTS & = & 0.045   _{-0.010}^{+0.022};     \\
\VTB & = & 0.77    _{-0.24}^{+0.18}.       
\end{eqnarray}
We test unitarity and determine that all six CKM normality 
conditions are consistent with unitarity. 
Four of these tests are performed for the first time.

\section*{Acknowledgements}
We thank the National Science Foundation for financial support
and the Department of Physics, Universidad Nacional de La Plata for 
their generous hospitality.



\begin{references}
\bibitem{CLEO_VUB_EXCL}
{J.P. Alexander \etal}, {CLEO CONF {\bf{96-16}} (unpublished); we do not use
  the updated result [CLEO CONF {\bf{97-30}} (unpublished)] of
  ${\mathrm{BR}}\left(B^0 \rightarrow \rho^-\ell^+\nu\right) = \left(2.0 \pm
  0.3 \pm 0.3\right)\times 10^{-4}$ since the uncertainty from the model
  dependence has yet to be evaluated}.
\end{references}
%
%

%
\renewcommand{\arraystretch}{1.15}
\begin{table}[!tb]
\caption{Results of the unitarity-free and unitarity-constrained fits described in the text.
         Values marked by a dagger ($\dagger$) are not explicitly free in the 
         fit but are derived from the other parameters in the same column.}
\label{tab:results}   
{\setlength{\tabcolsep}{0.01em}
\begin{tabular}{cll}                                  
\multicolumn{1}{c}{~}                                          &
\multicolumn{1}{c}{Free fit}                                   &
\multicolumn{1}{c}{Constrained fit}                             \\ \hline 
%
%
$\mZ$ (GeV)                     & $  91.187   \pm 0.002            $ & $  91.187  \pm 0.002        $  \\
$\mt$ (GeV)                     & $ 174.2     \pm 5.7              $ & $ 172.4    \pm 5.3          $  \\
$\log_{10}(\mH/{\mathrm{GeV}})$ & $   2.15    _{-0.38}^{+0.30}     $ & $   2.03   _{-0.37}^{+0.30} $  \\
$\alpha_{\mathrm{s}}(m_Z)$      & $   0.1171  \pm 0.0025           $ & $   0.1188 \pm 0.0021       $  \\
$\alpha^{-1}(m_Z)$              & $ 128.913   \pm 0.092            $ & $ 128.904  \pm 0.092        $  \\
$\VUD$  			& $   0.9740  \pm 0.0005           $ & $ 0.9748 \pm 0.0003         $  $\dagger$  \\ 
$\VUS$  			& $   0.2205  \pm 0.0018           $ & $ 0.2230 \pm 0.0014         $  $\dagger$  \\ 
$\VUB$  			& $   0.00325 \pm 0.00058          $ & $ 0.0032 \pm 0.0006         $  $\dagger$  \\ 
$\VCD$  			& $   0.224   \pm 0.016            $ & $ 0.2228 \pm 0.0014         $  $\dagger$  \\ 
$\VCS$  			& $   1.01    \pm 0.18             $ & $ 0.9741 \pm 0.0003         $  $\dagger$  \\ 
$\VCB$  			& $   0.0401  \pm 0.0029           $ & $ 0.0397 \pm 0.0029         $  $\dagger$  \\ 
$\VTD$  			& $   0.0113  _{-0.0029}^{+0.0060} $ & $ 0.0093 \pm 0.0018         $  $\dagger$  \\ 
$\VTS$  			& $   0.045   _{-0.010}^{+0.022}   $ & $ 0.0387 \pm 0.0029         $  $\dagger$  \\ 
$\VTB$  			& $   0.77    _{-0.24}^{+0.18}     $ & $ 0.9992 \pm 0.0001         $  $\dagger$  \\ 
$s_{12}$		        & $   -                            $ & $ 0.2230 \pm 0.0014         $  \\
$s_{13}$		        & $   -                            $ & $ 0.0032 \pm 0.0006         $  \\
$s_{23}$		        & $   -                            $ & $ 0.0396 \pm 0.0030         $  \\
$\cos\delta_{13}$		& $   -                            $ & $-0.26^{+0.82}_{-0.74}      $  \\
%
\end{tabular}
}
\end{table}
%

\end{document}